\documentclass[%
 amsmath,amssymb,
  a4paper,
  english,
  superscriptaddress
]{revtex4-1}
\usepackage{comment}
\usepackage{graphicx}
\usepackage{dcolumn}
\usepackage{bm}

\usepackage{todonotes}
\begin{document}


\title{Inside the Echo Chamber: Disentangling network dynamics from polarization}

\author{Duilio Balsamo}
\email{duilio.balsamo@unito.it}
\affiliation{University of Turin, Italy}

\author{Valeria Gelardi}
\email{valeriagelardi@gmail.com}
\affiliation{Aix Marseille Univ, Universit\'e de Toulon, CNRS, CPT, Marseille, France}

\author{Chengyuan Han}
\email{ch.han@fz-juelich.de}
\affiliation{Forschungszentrum J\"ulich, Germany }
\affiliation{University of Cologne,  Germany }

\author{Daniele Rama}
\email{daniele.rama@studenti.unito.it}
\affiliation{ISI Foundation, Turin, Italy}

\author{Abhishek Samantray}
\email{abhishek.samantray@imtlucca.it}
\affiliation{IMT School for Advanced Studies Lucca, Italy}

\author{Claudia Zucca}
\email{claudia.zucca@glasgow.ac.uk}
\affiliation{University of Glasgow,  United Kingdom}

\author{Michele Starnini}
\email{michele.starnini@gmail.com}
\affiliation{ISI Foundation, Turin, Italy}

\date{\today}

\begin{abstract}
Echo chambers are defined by the simultaneous presence of opinion polarization with respect to a controversial topic and homophily, i.e. the preference of individuals to interact with like-minded peers.
While recent efforts have been devoted to detecting the presence of echo chambers in polarized debates on online social media, the dynamics leading to the emergence of these phenomena remain unclear. 
Here, we contribute to this endeavour by proposing novel metrics to single out the effect of the network dynamics from the opinion polarization. 
By using a Twitter data set collected during a controversial political debate in Brazil in 2016, we employ a temporal network approach to gauge the strength of the echo chamber effect over time. 
We define a measure of opinion coherence in the network showing how the echo chamber becomes weaker across the observed period.
The analysis of the hashtags diffusion in the network shows that this is due to the increase of social interactions between users with opposite opinions.  Finally, the analysis of the mutual entropy between the opinions expressed and received by the users permits to quantify the social contagion effect.
We find empirical evidence that the polarization of the users and the dynamics of their interactions may evolve independently.
Our findings may be of interest to the broad array of researchers studying the dynamics of echo chambers and polarization in online social networks.
\end{abstract}
\flushbottom

\maketitle

\section{Introduction}



Various disciplines including social science, economics, physics, and others address  issues concerning the dynamics of social interactions from different perspectives. 
Researchers tend to provide qualitative understanding about how such interactions shape  social behavior, and also to provide quantifiable descriptions of such behaviour. 
A lack of data concerning social phenomena characterized the last century, making it difficult to study such interactions. 
However, availability of new sources of data and introduction of new  computational approaches have  enabled a quantitative description of such phenomena. Several topics in social sciences about human behavior and their interactions in online environments are being studied  by employing various quantitative methods  \citep{bakshy2015exposure, bond201261, grabowicz2012social}.  
For instance, there are studies  on the analysis of the content exchanged online  \citep{bail2016combining, gloor2009web},  the effects of algorithms employed in platforms such as browsers on behavior  \citep{lazer2015rise}, the use of network analysis to analyze social dynamics of opinion formation by modeling the interactions between actors \citep{watts2007influentials}. 
Several scholars already employed them to analyze online platforms such as Twitter to explain interactions between people \citep{barbera2015birds, cha2010measuring, kwak2010twitter}, the extent to which communication affects and eventually reshape public opinion \citep{masum2012reputation}, and the dynamics of collective behavior in response to events that catalyze public opinion \citep{tremayne2014anatomy, gonzalez2013broadcasters}. \\

The debate about the appropriateness and effectiveness of current measures is still ongoing \citep{tufekci2014big}, and the floor is open for new methodological contributions \citep{weaver2018dynamic}. 
One of the most salient debate concerns the extent to which online socialization affects the way people form opinions online. 
\citet{colleoni2014echo} point out the existence of two groups of scholars that provide antithetical evidence on those effects. On the one side, the Internet is deemed to facilitates selective exposure to contents, \citep{festinger1962theory} leading users to reinforce their pre-existing beliefs, due to homophily.  On the other side, the Internet challenges traditional social boundaries by exposing users to alternative opinions, views, and sources that they would not have accessed otherwise, allowing them to enlarge the so-called public sphere \citep{habermas1989structural}. \\

We contribute to this current debate by introducing new metrics to unfold the dynamics of opinion formation that drives mass level segregation. Single actors observed at the level of the network of interactions on Twitter, express their opinions in accordance with their political values as a response to external inputs driven by social interactions. Mass behavior is understood as the summation of individual behaviors of actors. Relying on the concept of echo chambers, we observe polarized trends in the mass opinion as a response to social events. 

\section{The Dynamics of Echo Chamber Segregation}

The concept of echo chamber was not invented in the last few years.  Introduced by \citet{key1966responsible} for characterizing the segregation of the American electorate in the sixties as resilient to the exposure to opinions in disagreement with their party affiliations,  echo chambers describe any closed environments in which the exchange of opinions is restricted to a group of like-minded people. Embedded in consolidated social science theory such as homophily  \citep{lazarsfeld1954friendship}  in sociology, cognitive dissonance \citep{festinger1962theory} in psychology and selective exposure \citep{klapper1960effects} in communication theory,  the usage of this concept boomed after the web 2.0 revolution \citep{o2007web} to snapshot the patterns of behavior in a connected world. \\

An increasingly high number of scholars is now enquiring after the existence of the echo chamber effect on social media.  Some of them provide empirical evidence of echo chambers formation \citep{sasahara2019inevitability, vaccari2016echo, williams2015network}; other show scenarios moderately segregated \citep{guess2018avoiding, garrett2009echo} or where segregation is explained by other factors such as individual features rather than social interactions \citep{dubois2018echo, colleoni2014echo}. Moreover, \cite{barbera2015tweeting} contend that previous work might have overestimated ideological segregation and that echo chambers are dynamic processes that change over time.\\

A crucial and open debate also concerns how to define and measure echo chambers. \citet{cota2019quantifying} relate echo chambers to polarization in the political discourse and suggest how to quantify the extent of the segregation. Several other works rely on the concept of opinion polarization in a particular discourse \cite{fisher2013does, garcia2015ideological}  mostly relying on community detection algorithms \citep{pujol2009divide}. Besides, \citet{jasny2015empirical} suggest to employ exponential random graph model to measure the formation of echo chambers differentiating the echo from the chamber in a triadic closure. \\

Despite the size of the debate, the reliability of theoretical explanations already offered as much as the empirical evidence presented, more work needs to be done to understand the dynamics of opinion formation in response to online inputs characterized by a push and pull dimension.
Echo chambers phenomena cannot exclusively be explained as polarized opinions that drive the formation of segregated community. Clusters successfully show the crystallization of a specific social state, but they do not explain the dynamics that prompted the formation of the state itself. What does drive this process? Which behaviour characterizes the creation of those communities?  \\

We contribute to these challenges by introducing a novel metric of opinion coherence in the network of interactions, that models how the individual responses to social event evolve.
At the very heart of segregation, indeed, there is the opinion of people that, rather than being stable over time, can change as a response to the external inputs \citet{lodge2013rationalizing}.
The aggregate assessment of the individual opinion can explain the evolution of mass behavior over time and to what extent the masses enter and exit the echo chambers as a function of their opinion and the pattern of their socialization. Together with the assessment of opinion coherence, we introduce a measure of the randomness of the opinion expressed by each user at each point in time. We compute the measure using Shannon's entropy \cite{shannon1948mathematical, song2010limits}. 
Finally, we dig into the dynamics of opinion exchange by analyzing the hashtags since they are a natural classification offered by the Twitter environment. By checking how topics spread in the network over time, we can provide an accurate description of the evolution of the interactions.\\

The triangulation of our measures substantially advances the current understanding of the echo chamber phenomena by allowing a separate observation of the segregation and polarization over time. While it is widely accepted to assess the extent of segregation through the measurement of polarization, our measures focus on the disentanglement of these two effects. Even if polarized groups very often segregate and interact into echo chambers, it is also extremely popular to observe cross-chamber interactions. The two populations might not agree on the object of discussion, but not necessarily stop the communication. The presence of strong opinions does not automatically lead to homophily,  or disconnection due to dissonant content, nor selective exposure. Our measures enable higher precision in the definition and measurement of the echo chamber phenomenon. \\

We implemented our approach on data scraped from Twitter in the most salient 43 weeks capturing the debate around the impeachment of President Dilma Rousseff in Brazil. The fierce nature of the competitive relationships between political actors in this Federal Republic in the last few decades make it a very suitable case for observing patterns of opinion evolution and polarization online.  Due to these particular political circumstances, other works already focused on social segregation online in Brazil \cite{recuero2015hashtags, recuero2019using, cota2019quantifying}, and this makes it a perfect scenario to test our measures.
  
\section{Results}

In our data set, each tweet is characterized by a sentiment, or opinion, $x = \{ -1, 0, +1\}$, corresponding to a pro-impeachment ($-1$), neutral ($0$), or anti-impeachment ($+1$) sentiment, respectively. 
The neutral opinion means that a hashtag can be used in both a positive or negative contexts. 
A user $i$ is thus characterized by a time-ordered set of expressed opinions $\mathcal{X}_i = \{x_i(t_1), x_i(t_2), \ldots, x_i(t_n)\}$, the size of this set representing his/her \textit{activity} $ a_i \equiv |\mathcal{X}_i|$.
We represent the social interactions between users as a directed temporal network (\citet{holme2012temporal}), in which a link is drawn from node $i$ to node $j$ at time $t$ if at least a tweet was posted at time $t$ by user $i$ mentioning user $j$. 
Since in our data set, we consider only tweets containing mentions, each tweet represents both an opinion expressed and a social interaction from one user to another. 
We then aggregate the data by slicing the ordered data stream in temporal windows with a fixed length, $\Delta t=1$ week.  
For each time interval $t = 1,2, \ldots, T$, with $T=43$ the total number of weeks, one can define a network, in which the weight $w_{ij}(t)$ between nodes $i$ and $j$ represents the number of tweets sent by user $i$ mentioning user $j$, in the time interval $t$. Conversely, one can define the average opinion expressed by user $i$ in that time interval, $x_i(t)$. 

With this notation, the total number of interactions from user $i$ to user $j$ in the whole temporal sequence is defined as $w_{ij} = \sum_{t=1}^T w_{ij}(t)$, which corresponds to the weight between node $i$ and node $j$ in the static, aggregated network. 
The total number of interactions in the time interval $t$, instead, is given by $W(t) = \sum_{(i,j)\in \mathcal{I}(t)} w_{ij}(t)$, where $\mathcal{I}(t)$ is the set of users interacting in the time interval $t$. 
Conversely, the total number of interactions over the whole time sequence is given by $W =\sum_{t=1}^T \sum_{(i,j)\in \mathcal{I}(t)} w_{ij}(t) = \sum_{t=1}^T W(t)$. Note that, by definition, $W$ is also equivalent to the sum of the activity of each user, $W=\sum_i a_i$.

We now characterize the dynamics of both the temporal network and opinions in time. 
Figure 4 in SM provides a picture of the intensity of activity over time measured by the number of tweets and the number of active users. 
While in the first weeks we see an intense activity, this trend drops in the following weeks, characterizing the highly dynamic nature of the interactions.  
Conversely, the average degree of the network also decreases in time, as shown in figure 5 of SM.
However, figure 6  and figure 7 of the SM show that despite the non-stationary nature of the dynamics, the degree distribution tend to be fairly stable over time ( $\mu = 1.5$  and $\sigma = 0.14$ of the best fit slopes) and the opinion distribution conserves a bi-modal trend (i.e. users' opinions remain polarized over time). 
Figure 8 of SM focuses on the correlation between the opinion expressed by a user and their activity, in different time intervals. 
The figure shows that more active users tend to larger polarization, resembling other highly polarized debates on online social media \cite{bakshy2015exposure}.
One can see that this structural feature is preserved over different time intervals.


\subsection{Coherence}
To understand the interplay between opinion and network dynamics over time, we introduce here a novel measure of opinion coherence in the network of interactions,
\begin{equation}
    \label{coherence}
    C(t) = \frac{1}{W(t)} \sum_{(i,j)\in E(t)} \! w_{ij}(t) \; f(\left|x_i(t) - x_j(t)\right|)
\end{equation}
which depends on the interactions between users at time $t$ (through the term $w_{ij}(t)$), as well as on the difference between their opinions, as given by the function $f(\left|x_i(t) - x_j(t)\right|)$.
This metric
encodes the dependency of the coherence from the distance between opinions of users $i$ and $j$. We choose a decreasing function of the opinion distance $ \left|x_i(t) - x_j(t)\right|$, such as 
$f(x) =  e^{-\beta x}$, with $\beta=1$.
This means that the contribution of the opinion distance in the coherence exponentially decreases as the opinions of the two users are moving away from each other. 
Therefore, when users characterized by similar opinions (small opinion distance) interact to a larger extent, the value of the function is higher, indicating a more coherent system.

\begin{figure}[b]
    \centering
    \includegraphics[width=0.6\paperwidth]{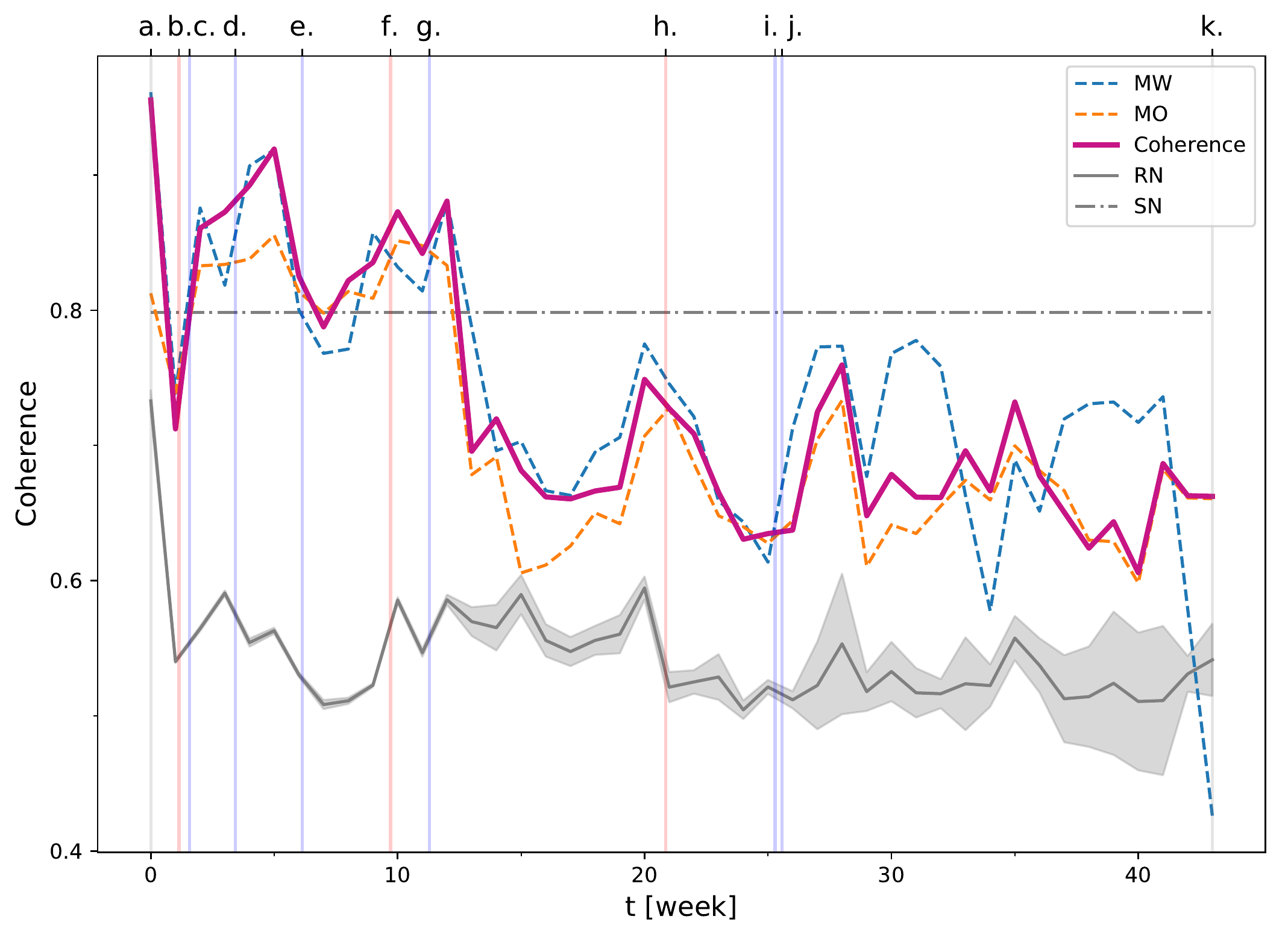}
    \caption{Evolution of opinion coherence in the network over time. We compare the empirical data (red curve) with several null models preserving different features of the data, see Methods for details. Vertical lines represent the polarization of the most important known events\footnote{a. Start collection (0), b. Biggest street demonstration against the government spread out in more than 250 (-1), c. Supreme court permits the constitution of a commission on the chamber of deputies (+1), d. MDB Brazilian political party “Movimento Democratico Brasileiro” left the government (+1), e. Deputy chamber approves impeachment with 367 votes against 137 (+1), f. Rousseff leaves the presidency after Senate approval (-1), g. Audio of Senator Romero Juca saying ”Estancar a sangria” (+1), h. Rousseff delivers final arguments in the Deputy chamber (-1), i. Rousseff’s defense in Senate (+1), j. Senate approves impeachment with 62 to 20 votes (+1), k. end of collection (0)}}
    \label{fig:coherence}
\end{figure}

Figure \ref{fig:coherence} shows the evolution of the coherence of the system for the entire period.
For comparison, we consider several null models, which preserve different features of the data.
Firstly, we consider a null model in which the opinions of the users are randomized within each time interval $t$ , resulting in assigning the opinion  $x_i(t)$ expressed by each user $i$  to a different user $j$.
This bootstrap approach corresponds to rewiring the interactions of the network and then recomputing their weights $w_{ij}(t)$. We indicate the network thus obtained  as  randomized network (RN).  In this case, the correlation between the interactions among the users  and the opinions expressed by them  are completely destroyed, thus dismantling the echo chamber.
A total of $100$ realizations are run, obtaining an ensemble of RNs. 
In Figure \ref{coherence} the timeline of the real coherence values is compared with the one of the RN. A significant difference can be observed between the true coherence and its typical bootstrapped realization, as the coherence from the data is far more than $1 \sigma$ away from the result of the RN null model. This suggests that the coherence measurement we constructed is a good indicator to quantify the echo chamber effect. 
We also consider other null models that preserve the interaction network but destroy other features of the data. 
The second one indicated as mean interaction (MW), substitutes the actual interactions at time $t$, $w_{ij}(t)$, by the average interactions $w_ij$ between user $i$ and user $j$ in the aggregated network, divided by $W$. 
A third null model, indicated as the mean opinion (MO),
substitutes the actual opinion $x_{i}(t)$ of user $i$ at time $t$  by his/her average opinion in the whole time sequence, $\bar{x}_i$. 
One can see that there is no significant difference between the coherence measured on empirical data and the results shown both with MW an MO null model, indicating respectively that the interactions between users and users-opinions are quite stationary throughout the evolution of discussion. Also, MW an MO result to be statistically significant with respect to the RN null model.  
This suggests that in a case in which the full information about the amount of interactions and opinions of the users are not accessible at each time representation of the network, the measure of coherence holds in its the mean opinion and mean interaction approaches.


\subsection{Hashtags}

\begin{figure}[b]
    \centering
    \includegraphics[width=0.6\paperwidth]{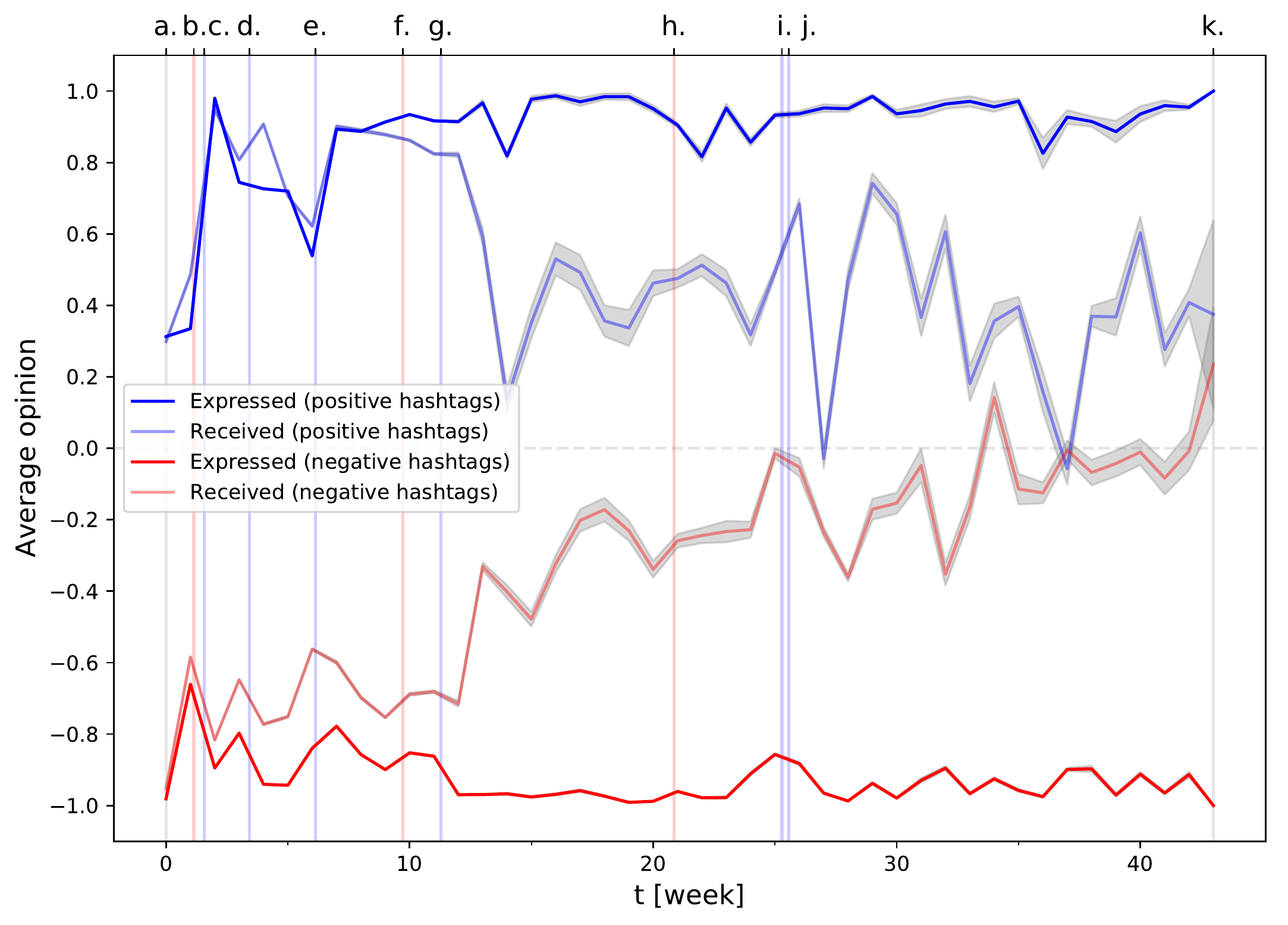}
    \caption{Average senders and receivers opinion for most relevant hashtags.}
    \label{fig:hashtags}
\end{figure}


The information provided by the coherence of the opinions in the network can be complemented by looking at the diffusion of the hashtags used in tweets by the users. 
We sampled the 20 most popular hashtags out of the all set.
Among these, we selected the most relevant hashtags according to the popularity and to the amount of time in which they were in use. 
The eight hashtags we select cover almost the entire time span. 
Hashtags are strongly associated with opinions: some of them convey a strongly pro-impeachment opinion (eg, ForaDilma, ForaPT), others convey a strongly anti-impeachment opinion (eg, ForaTemer, NaoVaiTerGolpe).
This is shown in Figure 9 in the SM, in which we plot the average opinion expressed by the users who tweeted using the selected hashtags, as a function of time.
In  the  figure,a strong polarization of the hashtag usage over entire period is observed, showing that users who share these hashtags are not inclined to opinion change.


Moreover, the hashtags' diffusion in the network can be used to measure the strength of the echo chamber.
One can measure the reciprocal exchange of information from one community to another, as the extent to which "positive" hashtags may reach users of different opinions, and vice versa. 
Figure \ref{fig:hashtags} shows, for each time interval, the average opinion of users sending the selected hashtags, as well as the average opinion of users who receive them, separately for hashtags associated with positive (blue) and negative (red) opinions.
While the average opinion of users sending the hashtags is constant in time and always very polarized, for both positive and negative hashtags, the average opinion of users receiving the hashtags becomes less and less extreme over time. 
This indicates that while at the beginning hashtags remained within the community using them, after some time they also reach users with different opinions, actually breaking the echo chambers.
This is true for both positive and negative echo chamber, as shown by Figure \ref{fig:hashtags}. 
These results are in agreement with the coherence dynamics (Figure \ref{fig:coherence}). High coherence corresponds to a small flow of information. As coherence decreases, the stream of interaction across parties intensifies.


\subsection{Mutual Information}

\begin{figure}[b]
    \centering
    \includegraphics[width=0.6\paperwidth]{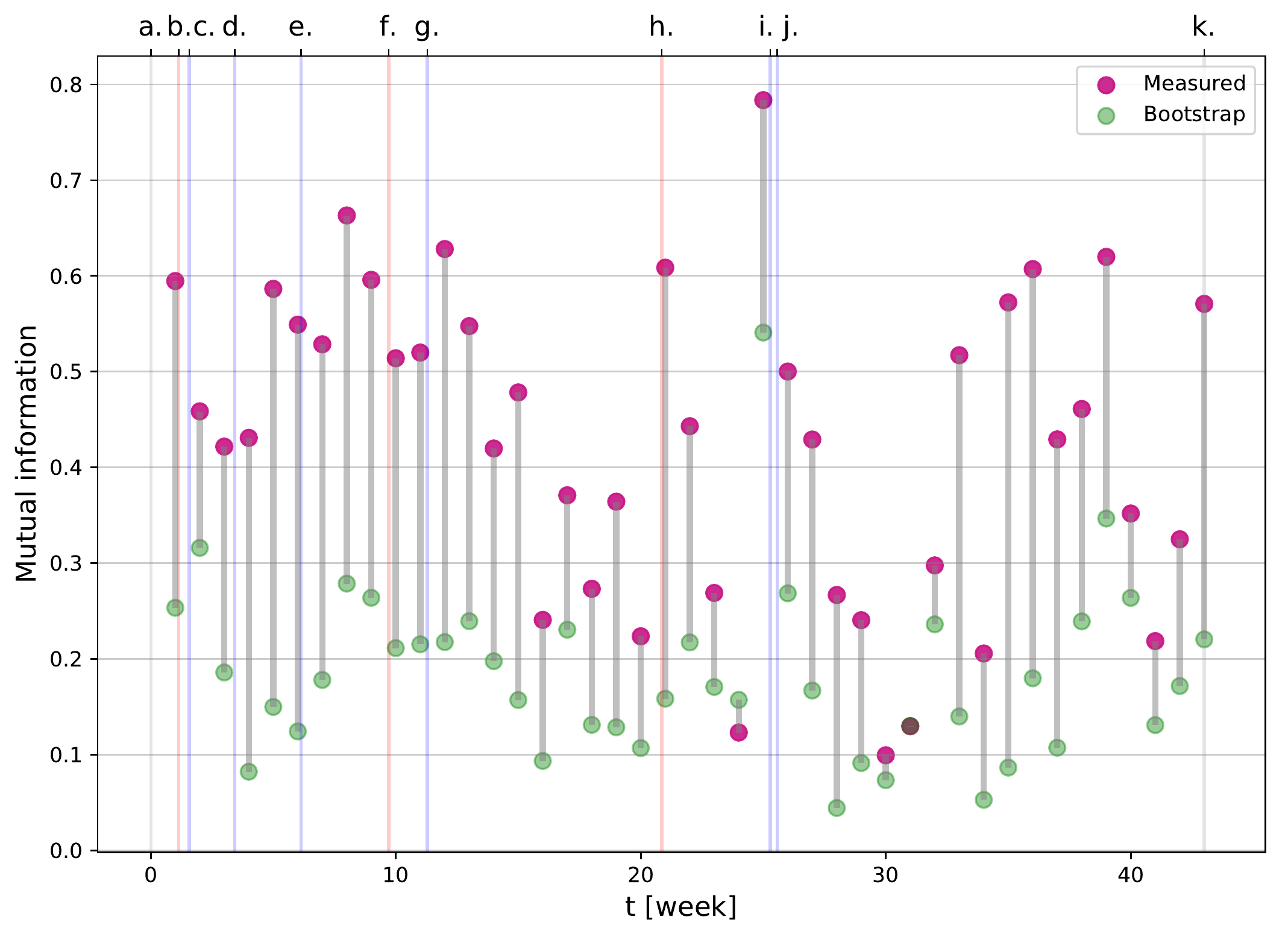}
    \caption{Difference between the Mutual Information evaluated at each time and evaluated with bootstrap.}
    \label{fig:mutual_information}
\end{figure}

Social contagion expresses the idea that individuals are able to influence the opinions of their peers \cite{christakis2013social}. 
In our case, tweets sent by users carry an opinion and are explicitly directed to other users, so can influence their opinions.
To understand the dynamics of this pattern we study the mutual information between the distribution of the opinion expressed by a set users, $p(x)$, and the distribution of the opinion they receive, $p(y)$, in each time interval. 

The mutual information in a time interval $t$ can thus be expressed as
\begin{equation}
    \label{mutual_info}
    MI(t) =  \sum_{x,y \in \mathcal{I}(t)} \! p_t(x,y) \ln \frac{p_t(x,y)}{p_t(x)p_t(y)}
\end{equation}
 where we select only those users which both express and receive and opinion in the time interval $t$, the set $\mathcal{I}(t)$,  and $p_t(x,y)$ is the joint distribution of the opinion expressed $x$ and received $y$.
 
This measure quantifies the degree of predictability between the opinions received by a user, and the opinions s/he is like to express, as in \citet{takaguchi2011predictability,starnini2017effects}. 
 A large value of $MI{(t)}$ suggests that the information contained in the distribution of expressed opinions is informative about the distribution of received opinions, and vice-versa. If the opinions expressed are independent by the opinions received, $p(x,y)=p(x)p(y)$, the mutual information is equal to zero. 
 Therefore, $MI(t)$ provides a quantitative understanding about the degree of deviation from  the received and expressed opinions being independent.

The mutual information in each time window is shown in Figure \ref{fig:mutual_information}. 
We compare the mutual information at each time interval to the bootstrapped mutual information estimated in a scenario in which network interactions are randomized, but the distributions of opinion expressed, $p(x)$, and received $p(y)$ are preserved.
Since mutual information is sensitive to sample size, and since the number of users that both express and receive and opinion, $\mathcal{I}(t)$ greatly varies among different time intervals $t$, the measured mutual information can not be compared between different time intervals. 
However, within each interval, we  see that the actual values are significantly higher compared to mean value of mutual information obtained after randomizing the interactions.  
This indicates that the opinions received by a user in an echo chamber environment play a key role in the predictability of the opinion of such user. 

\section{Discussion and Conclusion}
\label{Discussion}


This work contributes to the emerging literature in echo chambers and opinion formation by decoupling the effect of network interactions from the polarization of opinions in the system. 
In order to quantify the evolution of the echo chamber, we propose a novel metric of opinion coherence in the network, able to sense the strength of the echo chamber effect in time. 
The analysis of the hashtags diffusion over the network allows us to understand that the reason for the echo chamber effect to become weaker in time is to be found in the increase of the interactions between users with opposite opinions. 
Also, the analysis of the mutual entropy between the opinions expressed and received by the users permits to quantify the social contagion effect. \\

State of the art literature in psychology and political psychology concerning opinion formation do not conceive opinions as a skin that characterize people, but rather as a dress that changes in response to external inputs. This is due to a lack of rationality and cognitive coherence that characterize human behavior \citep{lodge2013rationalizing}. Our work relates to this consolidated theory by showing an analogous pattern in the echo chamber case. People do not live into echo chambers; they instead go in and out as a response to external inputs. 
Considering that the Mutual Information provides evidence of social contagion in the mechanisms that shape opinion formation, the received tweets are shown to affect the opinion formation over time. Hence, we can claim that social interaction is directly shaping the process of going in and out of the chamber. \\

The highly polarized opinion around Dilma Rousseff impeachment offered us a convenient opportunity to test our metrics. Future research should be devoted to applying  the metrics to other data sets of polarized debates in online social media, to quantify the echo chamber dynamics. 
For example, topics characterized by different degree of controversy are known to be different with respect to the strength of the echo chamber effect \cite{garimella2018political}.
Conversely, some social media, such as Facebook or Twitter, are known to be more prone to yield the conditions for the emergence of echo chambers than others, such as Reddit.

\section{Empirical Data}

     
\citet{cota2019quantifying} collected and presented the dataset we employed to test our measures. A team of Brazilian researchers scraped Twitter (from API) from March $5^{th}$ 2016 until December $31^{st}$ 2016 specifying a list of 323 keywords considered relevant to monitor the debate over former president Dilma Rousseff impeachment. The team selected only the tweets that contained at least one hashtag and a mention (retweets were discarded), preserving the timing of the interactions and the hashtags. The final dataset includes $2,066,042$ mentions and $285,670$ different users (people who wrote at least a tweet with a mention and an hashtag). 

A weighted temporal interaction network is computed from the data set, slicing the entire time-ordered data stream in time intervals of $N=7$ days and building an directed network for each time interval taking users as nodes, each mention from a user $i$ to another $j$ as directed link with weight equal to the number of time the user $i$ mentioned user $j$. 
For each time interval, an opinion value $x_{i}(t)$ is associated to the user $i$ as the average of the signs of the hashtags s/he sent in that time interval $t$.

\section{Acknowledgement}
This work is the output of the Complexity72h workshop, held at IMT School in Lucca, Italy, 17-21 June 2019. https://complexity72h.weebly.com/\\

We thank Wesley Cota,  Silvio C. Ferreira, and Romualdo Pastor-Satorras for sharing the dataset and making it available for this study.

\bibliographystyle{chicago}
\bibliography{references}

\begin{thebibliography}{}

\bibitem[\protect\citeauthoryear{Bail}{Bail}{2016}]{bail2016combining}
Bail, C.~A. (2016).
\newblock Combining natural language processing and network analysis to examine
  how advocacy organizations stimulate conversation on social media.
\newblock {\em Proceedings of the National Academy of Sciences\/}~{\em
  113\/}(42), 11823--11828.

\bibitem[\protect\citeauthoryear{Bakshy, Messing, and Adamic}{Bakshy
  et~al.}{2015}]{bakshy2015exposure}
Bakshy, E., S.~Messing, and L.~A. Adamic (2015).
\newblock Exposure to ideologically diverse news and opinion on facebook.
\newblock {\em Science\/}~{\em 348\/}(6239), 1130--1132.

\bibitem[\protect\citeauthoryear{Barber{\'a}}{Barber{\'a}}{2015}]{barbera2015birds}
Barber{\'a}, P. (2015).
\newblock Birds of the same feather tweet together: Bayesian ideal point
  estimation using twitter data.
\newblock {\em Political Analysis\/}~{\em 23\/}(1), 76--91.

\bibitem[\protect\citeauthoryear{Barber{\'a}, Jost, Nagler, Tucker, and
  Bonneau}{Barber{\'a} et~al.}{2015}]{barbera2015tweeting}
Barber{\'a}, P., J.~T. Jost, J.~Nagler, J.~A. Tucker, and R.~Bonneau (2015).
\newblock Tweeting from left to right: Is online political communication more
  than an echo chamber?
\newblock {\em Psychological science\/}~{\em 26\/}(10), 1531--1542.

\bibitem[\protect\citeauthoryear{Bond, Fariss, Jones, Kramer, Marlow, Settle,
  and Fowler}{Bond et~al.}{2012}]{bond201261}
Bond, R.~M., C.~J. Fariss, J.~J. Jones, A.~D. Kramer, C.~Marlow, J.~E. Settle,
  and J.~H. Fowler (2012).
\newblock A 61-million-person experiment in social influence and political
  mobilization.
\newblock {\em Nature\/}~{\em 489\/}(7415), 295.

\bibitem[\protect\citeauthoryear{Cha, Haddadi, Benevenuto, and Gummadi}{Cha
  et~al.}{2010}]{cha2010measuring}
Cha, M., H.~Haddadi, F.~Benevenuto, and K.~P. Gummadi (2010).
\newblock Measuring user influence in twitter: The million follower fallacy.
\newblock In {\em fourth international AAAI conference on weblogs and social
  media}.

\bibitem[\protect\citeauthoryear{Christakis and Fowler}{Christakis and
  Fowler}{2013}]{christakis2013social}
Christakis, N.~A. and J.~H. Fowler (2013).
\newblock Social contagion theory: examining dynamic social networks and human
  behavior.
\newblock {\em Statistics in medicine\/}~{\em 32\/}(4), 556--577.

\bibitem[\protect\citeauthoryear{Colleoni, Rozza, and Arvidsson}{Colleoni
  et~al.}{2014}]{colleoni2014echo}
Colleoni, E., A.~Rozza, and A.~Arvidsson (2014).
\newblock Echo chamber or public sphere? predicting political orientation and
  measuring political homophily in {T}witter using big data.
\newblock {\em Journal of Communication\/}~{\em 64\/}(2), 317--332.

\bibitem[\protect\citeauthoryear{Cota, Ferreira, Pastor-Satorras, and
  Starnini}{Cota et~al.}{2019}]{cota2019quantifying}
Cota, W., S.~C. Ferreira, R.~Pastor-Satorras, and M.~Starnini (2019).
\newblock Quantifying echo chamber effects in information spreading over
  political communication networks.
\newblock {\em arXiv preprint arXiv:1901.03688\/}.

\bibitem[\protect\citeauthoryear{Dubois and Blank}{Dubois and
  Blank}{2018}]{dubois2018echo}
Dubois, E. and G.~Blank (2018).
\newblock The echo chamber is overstated: the moderating effect of political
  interest and diverse media.
\newblock {\em Information, Communication \& Society\/}~{\em 21\/}(5),
  729--745.

\bibitem[\protect\citeauthoryear{Festinger}{Festinger}{1962}]{festinger1962theory}
Festinger, L. (1962).
\newblock {\em A Theory of Cognitive Dissonance}, Volume~2.
\newblock Redwood City, CA: Stanford University Press.

\bibitem[\protect\citeauthoryear{Fisher, Waggle, and Leifeld}{Fisher
  et~al.}{2013}]{fisher2013does}
Fisher, D.~R., J.~Waggle, and P.~Leifeld (2013).
\newblock Where does political polarization come from? locating polarization
  within the us climate change debate.
\newblock {\em American Behavioral Scientist\/}~{\em 57\/}(1), 70--92.

\bibitem[\protect\citeauthoryear{Garcia, Abisheva, Schweighofer, Serd{\"u}lt,
  and Schweitzer}{Garcia et~al.}{2015}]{garcia2015ideological}
Garcia, D., A.~Abisheva, S.~Schweighofer, U.~Serd{\"u}lt, and F.~Schweitzer
  (2015).
\newblock Ideological and temporal components of network polarization in online
  political participatory media.
\newblock {\em Policy \& Internet\/}~{\em 7\/}(1), 46--79.

\bibitem[\protect\citeauthoryear{Garimella, De~Francisci~Morales, Gionis, and
  Mathioudakis}{Garimella et~al.}{2018}]{garimella2018political}
Garimella, K., G.~De~Francisci~Morales, A.~Gionis, and M.~Mathioudakis (2018).
\newblock Political discourse on social media: Echo chambers, gatekeepers, and
  the price of bipartisanship.
\newblock In {\em Proceedings of the 2018 World Wide Web Conference on World
  Wide Web}, pp.\  913--922. International World Wide Web Conferences Steering
  Committee.

\bibitem[\protect\citeauthoryear{Garrett}{Garrett}{2009}]{garrett2009echo}
Garrett, R.~K. (2009).
\newblock Echo chambers online?: Politically motivated selective exposure among
  internet news users.
\newblock {\em Journal of Computer-Mediated Communication\/}~{\em 14\/}(2),
  265--285.

\bibitem[\protect\citeauthoryear{Gloor, Krauss, Nann, Fischbach, and
  Schoder}{Gloor et~al.}{2009}]{gloor2009web}
Gloor, P.~A., J.~Krauss, S.~Nann, K.~Fischbach, and D.~Schoder (2009).
\newblock Web science 2.0: Identifying trends through semantic social network
  analysis.
\newblock In {\em 2009 International Conference on Computational Science and
  Engineering}, Volume~4, pp.\  215--222. IEEE.

\bibitem[\protect\citeauthoryear{Gonz{\'a}lez-Bail{\'o}n, Borge-Holthoefer, and
  Moreno}{Gonz{\'a}lez-Bail{\'o}n et~al.}{2013}]{gonzalez2013broadcasters}
Gonz{\'a}lez-Bail{\'o}n, S., J.~Borge-Holthoefer, and Y.~Moreno (2013).
\newblock Broadcasters and hidden influentials in online protest diffusion.
\newblock {\em American Behavioral Scientist\/}~{\em 57\/}(7), 943--965.

\bibitem[\protect\citeauthoryear{Grabowicz, Ramasco, Moro, Pujol, and
  Eguiluz}{Grabowicz et~al.}{2012}]{grabowicz2012social}
Grabowicz, P.~A., J.~J. Ramasco, E.~Moro, J.~M. Pujol, and V.~M. Eguiluz
  (2012).
\newblock Social features of online networks: The strength of intermediary ties
  in online social media.
\newblock {\em PloS one\/}~{\em 7\/}(1), e29358.

\bibitem[\protect\citeauthoryear{Guess, Lyons, Nyhan, and Reifler}{Guess
  et~al.}{2018}]{guess2018avoiding}
Guess, A., B.~Lyons, B.~Nyhan, and J.~Reifler (2018).
\newblock Avoiding the echo chamber about echo chambers: Why selective exposure
  to like-minded political news is less prevalent than you think.
\newblock {\em Documento de la Knight Foundation. En l{\'\i}nea:
  https://kf-site-production. s3. amazonaws.
  com/media\_elements/files/000/000/133/original/Topos\_KF\_
  White-Paper\_Nyhan\_V1. pdf\/}.

\bibitem[\protect\citeauthoryear{Habermas}{Habermas}{1989}]{habermas1989structural}
Habermas, J. (1989).
\newblock The structural transformation of the public sphere, trans. thomas
  burger.
\newblock {\em Cambridge: MIT Press\/}~{\em 85}, 85--92.

\bibitem[\protect\citeauthoryear{Holme and Saram{\"a}ki}{Holme and
  Saram{\"a}ki}{2012}]{holme2012temporal}
Holme, P. and J.~Saram{\"a}ki (2012).
\newblock Temporal networks.
\newblock {\em Physics reports\/}~{\em 519\/}(3), 97--125.

\bibitem[\protect\citeauthoryear{Jasny, Waggle, and Fisher}{Jasny
  et~al.}{2015}]{jasny2015empirical}
Jasny, L., J.~Waggle, and D.~R. Fisher (2015).
\newblock An empirical examination of echo chambers in us climate policy
  networks.
\newblock {\em Nature Climate Change\/}~{\em 5\/}(8), 782.

\bibitem[\protect\citeauthoryear{Key}{Key}{1966}]{key1966responsible}
Key, V.~O. (1966).
\newblock {\em The Responsible Electorate}.
\newblock Cambridge, MA: Belknap Press of Harvard University Press.

\bibitem[\protect\citeauthoryear{Klapper}{Klapper}{1960}]{klapper1960effects}
Klapper, J.~T. (1960).
\newblock {\em The Effects of Mass Communication}.
\newblock New York: Free Press.

\bibitem[\protect\citeauthoryear{Kwak, Lee, Park, and Moon}{Kwak
  et~al.}{2010}]{kwak2010twitter}
Kwak, H., C.~Lee, H.~Park, and S.~Moon (2010).
\newblock What is twitter, a social network or a news media?
\newblock In {\em Proceedings of the 19th international conference on World
  wide web}, pp.\  591--600. AcM.

\bibitem[\protect\citeauthoryear{Lazarsfeld and Merton}{Lazarsfeld and
  Merton}{1954}]{lazarsfeld1954friendship}
Lazarsfeld, P.~F. and R.~Merton (1954).
\newblock Friendship as a social process: A substantive and methodological
  analysis.
\newblock In M.~Berger (Ed.), {\em Freedom and Control in Modern Society}, pp.\
   8--66. New York: Van Nostrand.

\bibitem[\protect\citeauthoryear{Lazer}{Lazer}{2015}]{lazer2015rise}
Lazer, D. (2015).
\newblock The rise of the social algorithm.
\newblock {\em Science\/}~{\em 348\/}(6239), 1090--1091.

\bibitem[\protect\citeauthoryear{Lodge and Taber}{Lodge and
  Taber}{2013}]{lodge2013rationalizing}
Lodge, M. and C.~S. Taber (2013).
\newblock {\em The Rationalizing Voter}.
\newblock New York: Cambridge University Press.

\bibitem[\protect\citeauthoryear{Masum and Tovey}{Masum and
  Tovey}{2012}]{masum2012reputation}
Masum, H. and M.~Tovey (2012).
\newblock {\em The reputation society: How online opinions are reshaping the
  offline world (The Information Society Series)}.
\newblock The MIT Press.

\bibitem[\protect\citeauthoryear{O'Reilly}{O'Reilly}{2005}]{o2007web}
O'Reilly, T. (2005, September).
\newblock {What Is Web 2.0? Design Patterns and Business Models for the Next
  Generation of Software}.
\newblock www.oreilly.com.

\bibitem[\protect\citeauthoryear{Pujol, Erramilli, and Rodriguez}{Pujol
  et~al.}{2009}]{pujol2009divide}
Pujol, J.~M., V.~Erramilli, and P.~Rodriguez (2009).
\newblock Divide and conquer: Partitioning online social networks.
\newblock {\em arXiv preprint arXiv:0905.4918\/}.

\bibitem[\protect\citeauthoryear{Recuero, Zago, Bastos, and Ara{\'u}jo}{Recuero
  et~al.}{2015}]{recuero2015hashtags}
Recuero, R., G.~Zago, M.~T. Bastos, and R.~Ara{\'u}jo (2015).
\newblock Hashtags functions in the protests across brazil.
\newblock {\em SAGE Open\/}~{\em 5\/}(2), 2158244015586000.

\bibitem[\protect\citeauthoryear{Recuero, Zago, and Soares}{Recuero
  et~al.}{2019}]{recuero2019using}
Recuero, R., G.~Zago, and F.~Soares (2019).
\newblock Using social network analysis and social capital to identify user
  roles on polarized political conversations on twitter.
\newblock {\em Social Media+ Society\/}~{\em 5\/}(2), 2056305119848745.

\bibitem[\protect\citeauthoryear{Sasahara, Chen, Peng, Ciampaglia, Flammini,
  and Menczer}{Sasahara et~al.}{2019}]{sasahara2019inevitability}
Sasahara, K., W.~Chen, H.~Peng, G.~L. Ciampaglia, A.~Flammini, and F.~Menczer
  (2019).
\newblock On the inevitability of online echo chambers.
\newblock {\em arXiv preprint arXiv:1905.03919\/}.

\bibitem[\protect\citeauthoryear{Shannon}{Shannon}{1948}]{shannon1948mathematical}
Shannon, C.~E. (1948).
\newblock A mathematical theory of communication.
\newblock {\em The Bell System Technical Journal\/}~{\em 27}, 379–--423.

\bibitem[\protect\citeauthoryear{Song, Qu, Blumm, and Barab{\'a}si}{Song
  et~al.}{2010}]{song2010limits}
Song, C., Z.~Qu, N.~Blumm, and A.-L. Barab{\'a}si (2010).
\newblock Limits of predictability in human mobility.
\newblock {\em Science\/}~{\em 327\/}(5968), 1018--1021.

\bibitem[\protect\citeauthoryear{Starnini, Baronchelli, and
  Pastor-Satorras}{Starnini et~al.}{2017}]{starnini2017effects}
Starnini, M., A.~Baronchelli, and R.~Pastor-Satorras (2017).
\newblock Effects of temporal correlations in social multiplex networks.
\newblock {\em Scientific reports\/}~{\em 7\/}(1), 8597.

\bibitem[\protect\citeauthoryear{Takaguchi, Nakamura, Sato, Yano, and
  Masuda}{Takaguchi et~al.}{2011}]{takaguchi2011predictability}
Takaguchi, T., M.~Nakamura, N.~Sato, K.~Yano, and N.~Masuda (2011).
\newblock Predictability of conversation partners.
\newblock {\em Physical Review X\/}~{\em 1\/}(1), 011008.

\bibitem[\protect\citeauthoryear{Tremayne}{Tremayne}{2014}]{tremayne2014anatomy}
Tremayne, M. (2014).
\newblock Anatomy of protest in the digital era: A network analysis of twitter
  and occupy wall street.
\newblock {\em Social Movement Studies\/}~{\em 13\/}(1), 110--126.

\bibitem[\protect\citeauthoryear{Tufekci}{Tufekci}{2014}]{tufekci2014big}
Tufekci, Z. (2014).
\newblock Big questions for social media big data: Representativeness, validity
  and other methodological pitfalls.
\newblock In {\em Eighth International AAAI Conference on Weblogs and Social
  Media}.

\bibitem[\protect\citeauthoryear{Vaccari, Valeriani, Barber{\'a}, Jost, Nagler,
  and Tucker}{Vaccari et~al.}{2016}]{vaccari2016echo}
Vaccari, C., A.~Valeriani, P.~Barber{\'a}, J.~T. Jost, J.~Nagler, and J.~A.
  Tucker (2016).
\newblock Of echo chambers and contrarian clubs: Exposure to political
  disagreement among german and italian users of twitter.
\newblock {\em Social Media+ Society\/}~{\em 2\/}(3), 2056305116664221.

\bibitem[\protect\citeauthoryear{Watts and Dodds}{Watts and
  Dodds}{2007}]{watts2007influentials}
Watts, D.~J. and P.~S. Dodds (2007).
\newblock Influentials, networks, and public opinion formation.
\newblock {\em Journal of consumer research\/}~{\em 34\/}(4), 441--458.

\bibitem[\protect\citeauthoryear{Weaver, Williams, Cioroianu, Williams, Coan,
  and Banducci}{Weaver et~al.}{2018}]{weaver2018dynamic}
Weaver, I.~S., H.~Williams, I.~Cioroianu, M.~Williams, T.~Coan, and S.~Banducci
  (2018).
\newblock Dynamic social media affiliations among uk politicians.
\newblock {\em Social networks\/}~{\em 54}, 132--144.

\bibitem[\protect\citeauthoryear{Williams, McMurray, Kurz, and
  Lambert}{Williams et~al.}{2015}]{williams2015network}
Williams, H.~T., J.~R. McMurray, T.~Kurz, and F.~H. Lambert (2015).
\newblock Network analysis reveals open forums and echo chambers in social
  media discussions of climate change.
\newblock {\em Global Environmental Change\/}~{\em 32}, 126--138.

\end{thebibliography}

\newpage

\Large\centering{Supplementary material} \\
\centering{Inside the Echo Chamber: Disentangling network dynamics from polarization}


\begin{figure}[h!]
    \centering
    \includegraphics[width=0.6\textwidth]{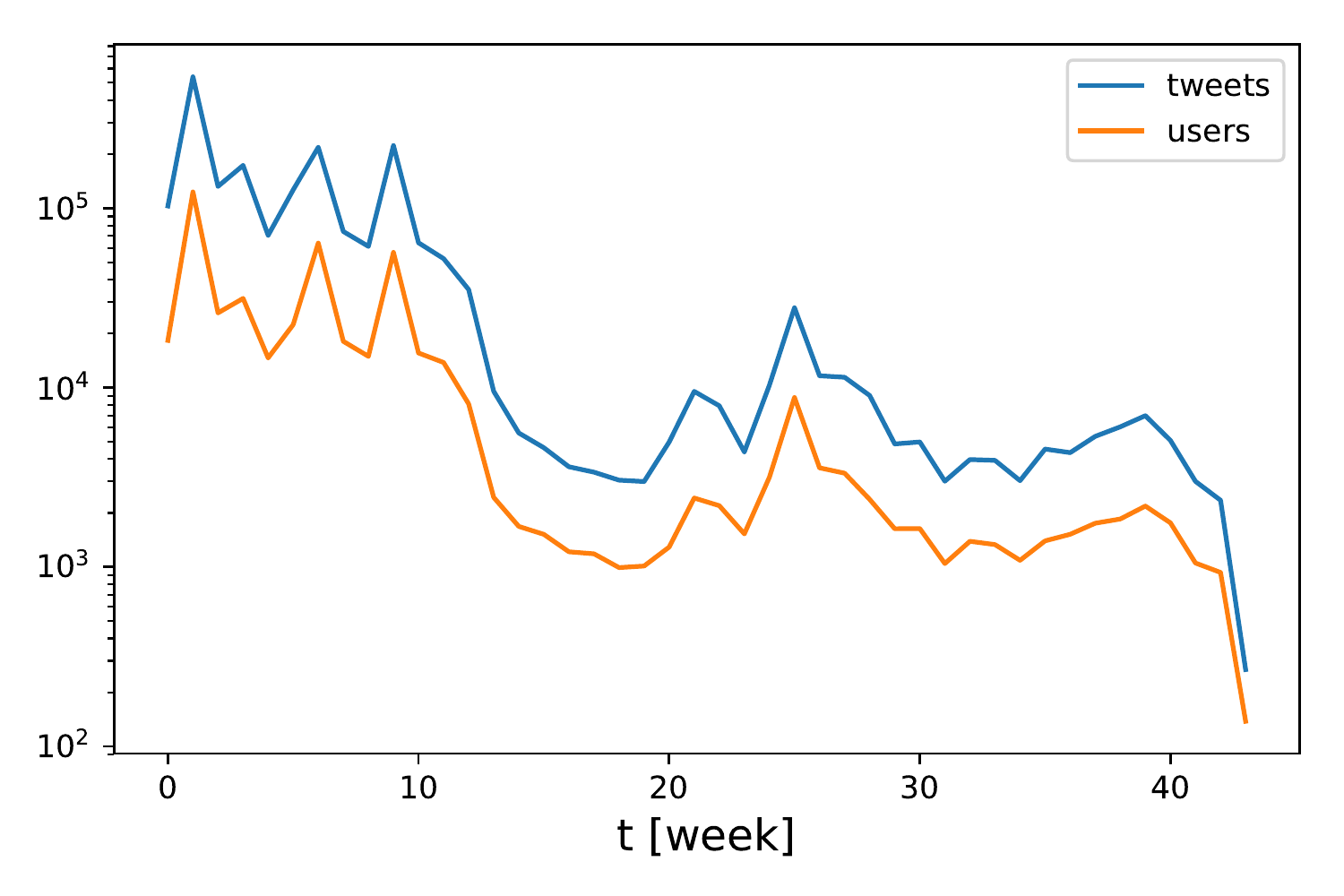}
    \caption{Number of tweets (in blue) and of users (in orange) in time. The first 9 weeks are characterized by high activity and high numbers of users, with a peak of more than half-million tweets per week, whereas in the successive weeks both the number of tweets and the number of users drastically drops.}
    \label{fig:tweets_users_vs_t}
\end{figure}

\begin{figure}[h!]
    \centering
    \includegraphics[width=0.6\textwidth]{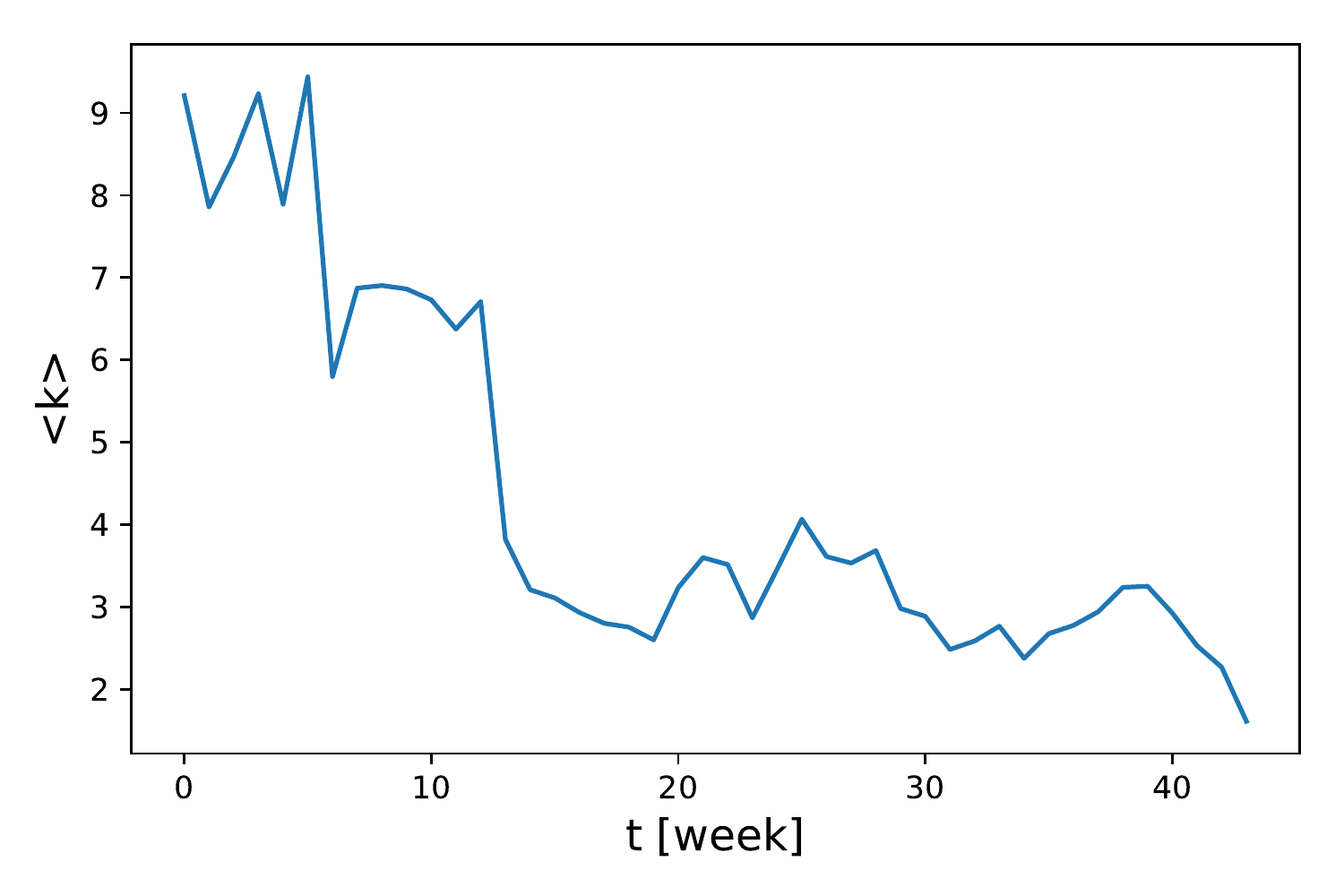}
    \caption{Average degree of the interaction network in time. The figure shows that the average degree of the network slightly decrease over time. }
    \label{fig:avg_deg_vs_t}
\end{figure}

\begin{figure}
    \centering
    \includegraphics[width=0.7\textwidth]{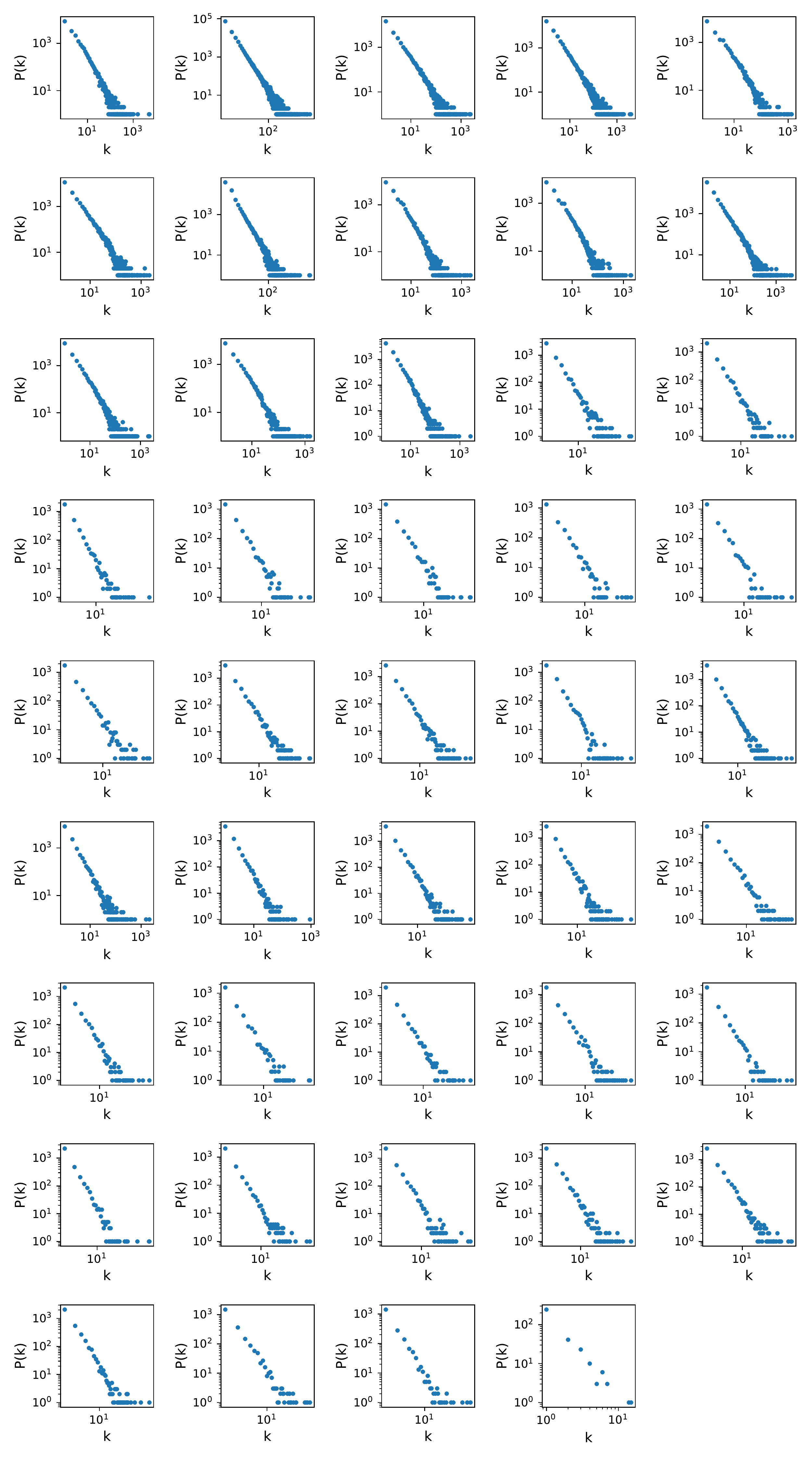}
    \caption{Degree distributions for each time window.They appear as heavy-tailed and they are consistent over time.}
    \label{fig:p(k)_vs_k}
\end{figure}

\begin{figure}[h!]
    \centering
    \includegraphics[width=0.7\textwidth]{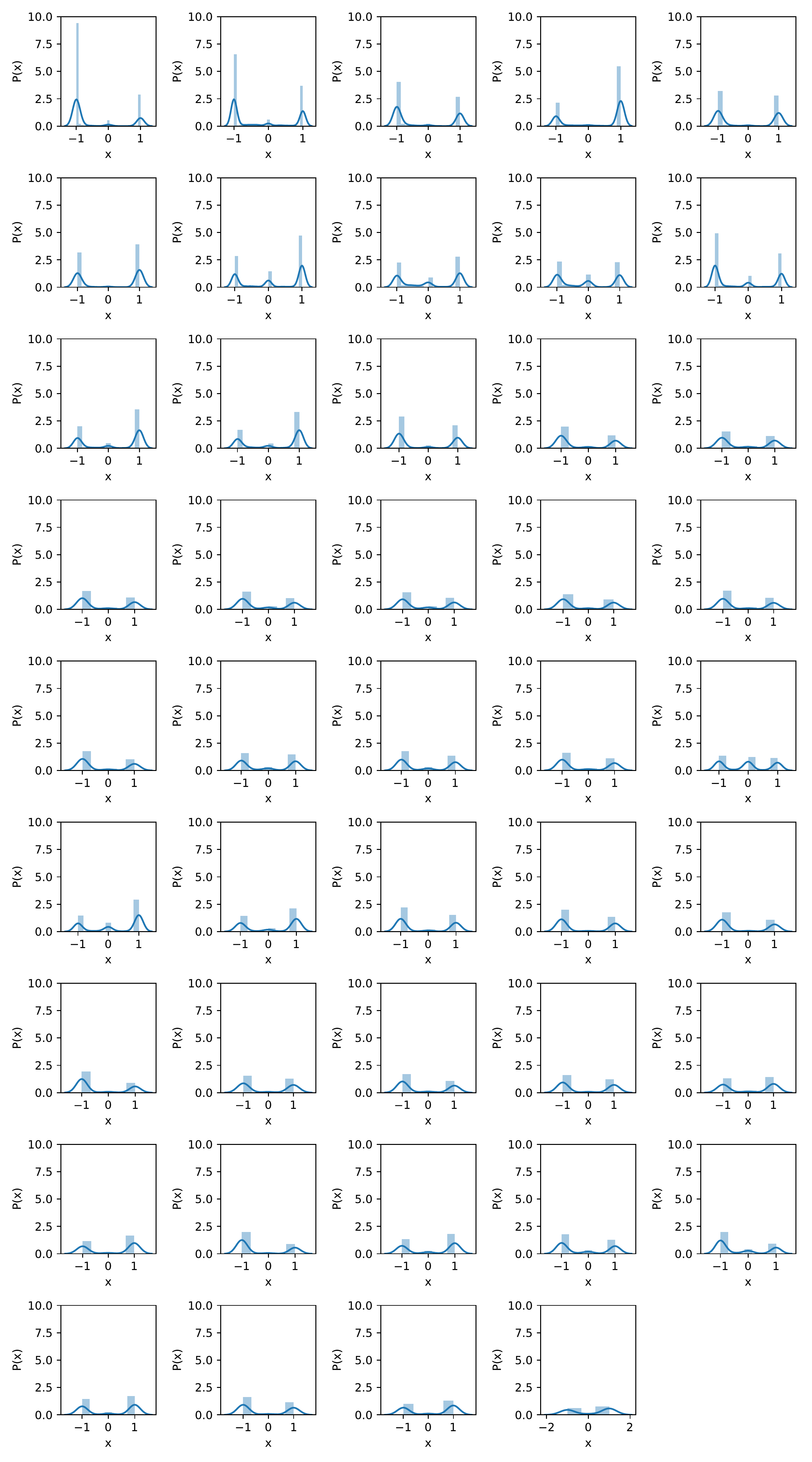}
    \caption{Distributions of opinions across users for each time window (on the y-axis a density measure is indicated). Distributions are bi-modal across all the time windows, showing a persistence of strong  polarization of users' opinions through time.}
    \label{fig:p(x)_vs_x}
\end{figure}

\begin{figure}[h!]
    \centering
    \includegraphics[width=0.7\textwidth]{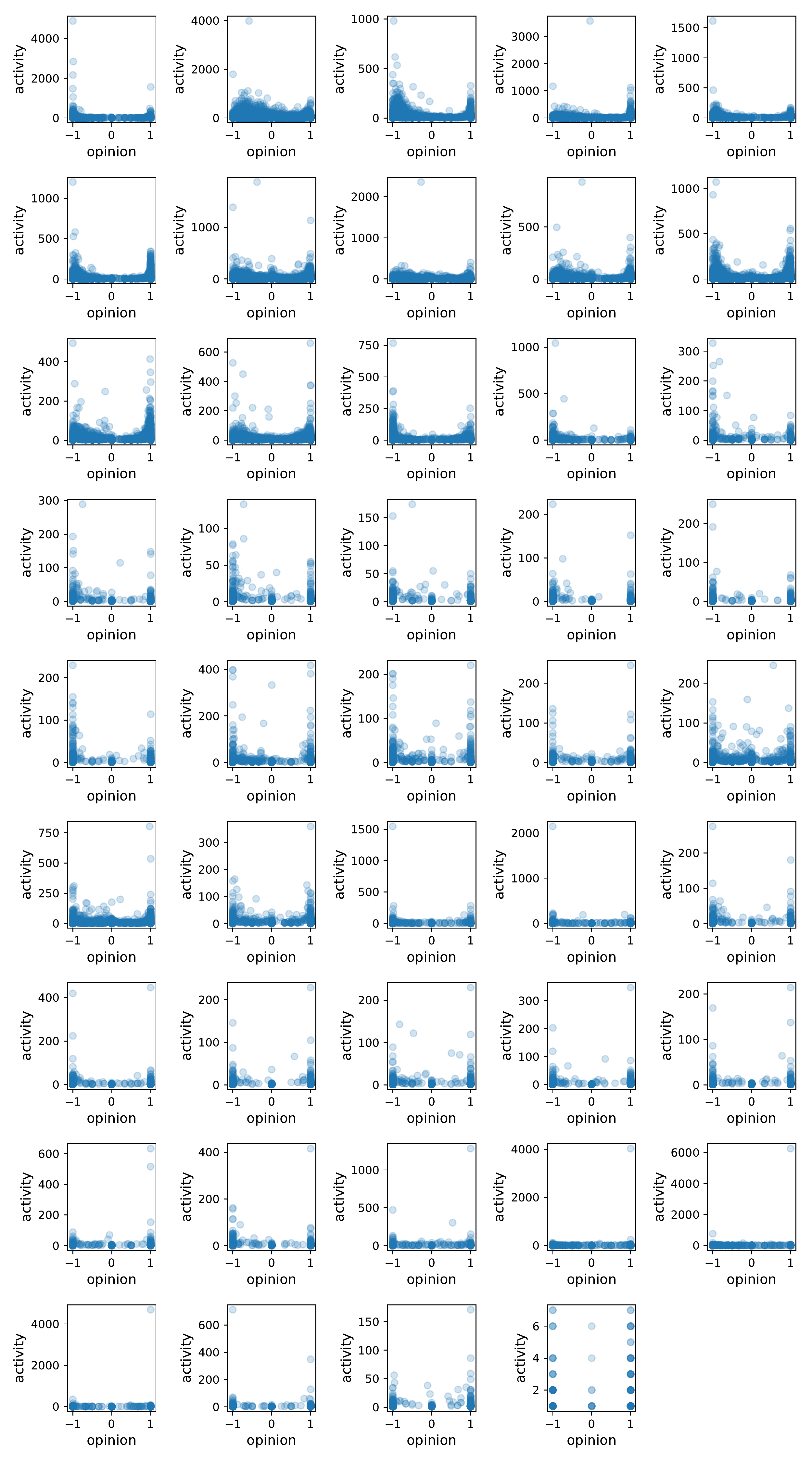}
    \caption{Activity vs. opinion for every user (blue points) for each time window. The higher density in correspondence of extreme values of opinion (either close to $-1$ or to $+1$), reveals again a polarization effect. Moreover, most active users revealed to be those with more polarized opinions.}
    \label{fig:activity_vs_x}
\end{figure}

\begin{figure}[h!]
    \centering
    \includegraphics[width=0.6\paperwidth]{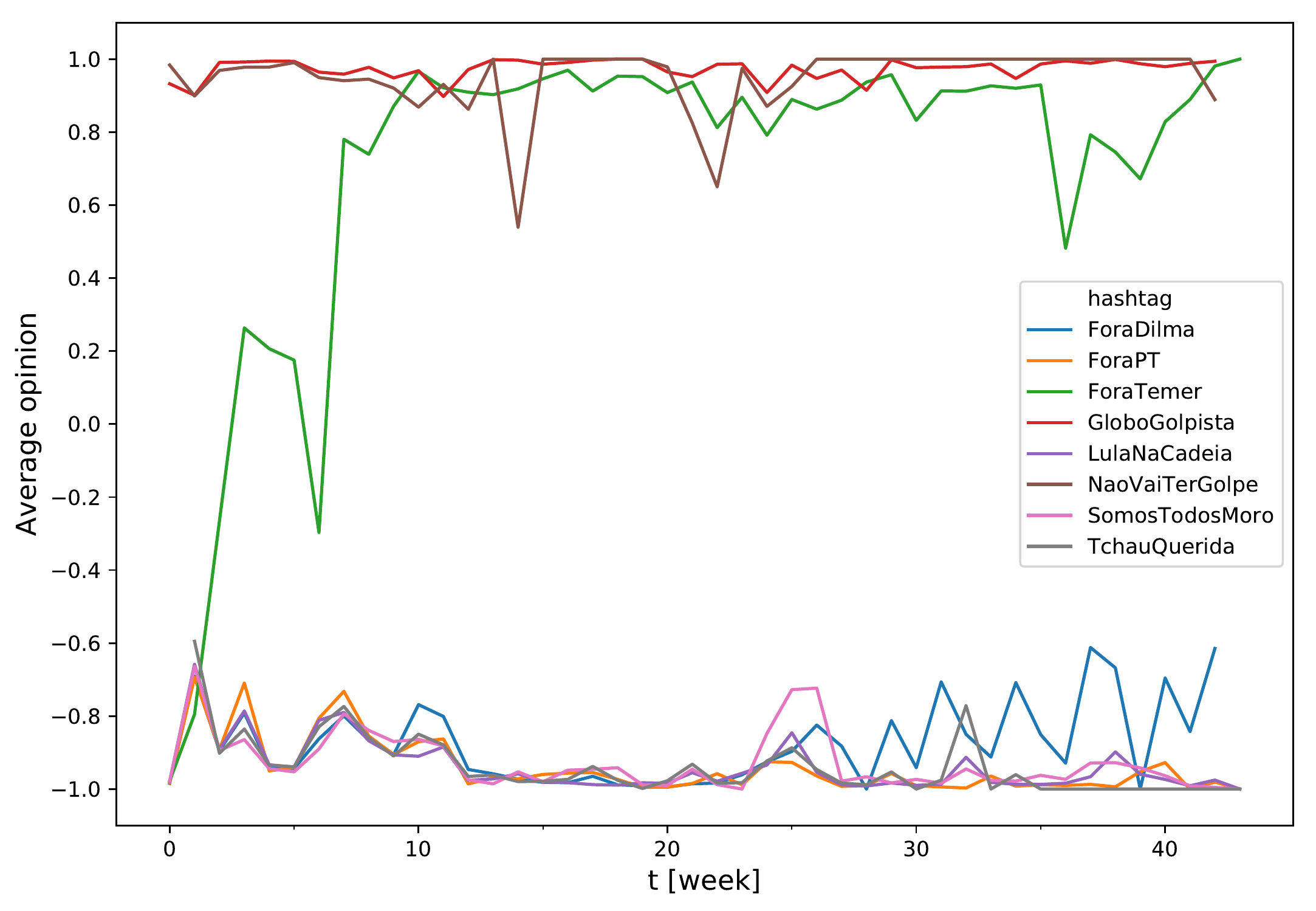}
    \caption{ Average opinion of the users that use the selected hashtags over time. }
    \label{fig:hashtags_selection}
\end{figure}

\end{document}